\documentclass{article}
\usepackage{epsfig}

\newcommand{\sheptitle}
{The flavour problem and family symmetry}

\newcommand{\shepauthor}
{S. F. King\footnote{Talk given by S.F.King}  and I. N. R. Peddie}

\newcommand{\shepaddress}
{
School of Physics and Astronomy,\\ 
University of Southampton, Southampton, SO17 1BJ, U.K.\\ 
E-mail: king@soton.ac.uk, inrp@soton.ac.uk
}

\newcommand{\shepabstract}
{
We show how two different family symmetries can be used to address
the flavour problem in $SO(10)$-like models. The first is based
on a gauged $U(1)_F$, whose problems dissappear  in the context
of a type I string model embedding. The second is based on 
$SU(3)_F$, the maximal family group consistent 
with $SO(10)$; in this case the family symmetry is more constaining,
so we merely look at it in the context of a supersymmetric field theory
}

\begin{document}

\begin{titlepage}
\begin{flushright}
  hep-ph/0309237 \\
  SHEP/0329
\end{flushright}
\begin{center}
  {\large{\bf \sheptitle}}
  \\ \shepauthor \\ \mbox{} \\ {\it \shepaddress} \\
  {\bf Abstract} \\ \bigskip
\end{center}
\setcounter{page}{0}
\shepabstract
\begin{flushleft}
\today
\end{flushleft}
\end{titlepage}

\newpage
\section{Introduction}

The flavour problem in SUSY models falls into two parts. The first
is to understand the origin of the Yukawa couplings, and with the 
see-saw mechanism to understand the origin of the Majorana 
masses. The second is to understand why flavour changing and CP violating
processes induced by SUSY loops is so small. Any theory of flavour should
address both problems simultaneously. Consider, $\mu\rightarrow e\gamma$
\cite{taumugamma}
:

\begin{equation}
  \label{eq:1}
  BR(\mu\rightarrow e\gamma) \approx
  \frac{\alpha^3}{G_F^2}f_{21}(M_2,\mu,m_{\tilde\nu})
  \left|
    m^2_{\tilde{L}_{21}}
  \right|^2 \tan^2\beta
\end{equation}

The simplest way to suppress processes like these is to suppress off
diagonal sfermion masses. There are two general ways of generating off
diagonal sfermion masses. The first, `primordial', is where there are
off diagonal elements in the SCKM basis.
 The second is RGE generated, where either Higgs triplets in running from 
the GUT scale to the EW scale, or for leptons, running the MSSM with right
handed neutrinos from the GUT scale to the lightest RH neutrino mass. In
general both will be present; models of flavour are concerned with suppressing
primordial contributions.

\section{$U(1)$ Family symmetry}

We consider here the $SO(10)$-like model of flavour 
$SU(4)_{PS} \otimes SU(2)_L \otimes SU(2)_R \otimes U(1)_F$. Here the
family index is promoted to being a gauge index under $U(1)_F$, and the
matter representations under $4,2,2,1$ are:

\begin{equation}
  \label{eq:2}
  F^i = (4,2,1)_{(1,0,0)}\;;\;
  \overline{F}^i = (\overline{4},1,\overline{2})_{(4,2,0)}
\end{equation}
\begin{equation}
  \label{eq:3}
  h = (1,\overline{2},2)_0
\end{equation}

The exotic Higgs fields are $H$,$\overline{H}$, and $\Phi$. The first
two break $SU(4)_{PS}\otimes SU(2)_R \rightarrow SU(3)_c \otimes U(1)_Y$.
The latter breaks $U(1)_F$. Their representation and VEVs are:

\begin{equation}
  \label{eq:4}
  H = (4,1,2)_0\;,\; \frac{\left<H\right>}{M} \approx \sqrt{\delta}
  \;;\;
  \overline{H} = (\overline{4},1,\overline{2})_0\;,\;
  \frac{\left<\overline{H}\right>}{M} \approx \sqrt{\delta}
\end{equation}
\begin{equation}
  \label{eq:5}
  \Phi = (1,1,1)_{-1}\;,\; \frac{\left<\Phi\right>}{M} \approx \epsilon
\end{equation}

The phenomenological values of the expansion parameters are 
$\delta = \epsilon = 0.22$. These are of use when we use Froggatt-Nielsen
operators \cite{Froggatt:1978nt} to generate the Yukawa matrices
 and the Majorana matrix for the RH neutrino fields:

\begin{equation}
  \label{eq:6}
  \mathrm{Yukawa:}\;\;
  F^i \overline{F}^j 
  \left(
    \frac{H\overline{H}}{M^2}
  \right)^{n_{ij}}
  \left(
    \frac{\Phi}{M}
    \right)^{p_{ij}}
    \rightarrow F^i \overline{F}^j C_{ij} \delta^{n_ij}\epsilon^{p_ij}
\end{equation}
\begin{equation}
  \label{eq:8}
  \mathrm{Majorana:}\;\;
  \overline{F}^i\overline{F}^j
  \left(
    \frac{HH}{M}
  \right)
  \left(
    \frac{H\overline{H}}{M^2}
  \right)
  \left(
    \frac{\Phi}{M}
  \right)^{q_{ij}}
  \rightarrow
  \overline{F}^i\overline{F}^j M_3 \delta \epsilon^{q_{ij}}
\end{equation}

With a certain operator texture \cite{Blazek:2003wz}, we can then write out
the Majorana matrix, and the Yukawa matrices, at the GUT scale:

\begin{equation}
  \label{eq:7}
  \frac{M_{RR}(M_X)}{M_3} \sim
  \left(
    \begin{array}{ccc}
      \delta\epsilon^8 & \delta\epsilon^6 & \delta\epsilon^4 \\
      \delta\epsilon^6 & \delta\epsilon^4 & \delta\epsilon^2 \\
      \delta\epsilon^4 & \delta\epsilon^2 & 1
    \end{array}
  \right)
\end{equation}

\begin{equation}
  \label{eq:9}
  Y^u(M_X) \sim 
  \left(
    \begin{array}{ccc}
      \sqrt{2}\delta^3\epsilon^5 & 
      \sqrt{2}\delta^2\epsilon^3 &
      \frac{2}{\sqrt{5}}\delta^2\epsilon \\
      0 &
      \frac{8}{5\sqrt{5}}\delta^2\epsilon^2 &
      0 \\
      0 &
      \frac{8}{5}\delta^2\epsilon^2 &
      r_t
    \end{array}
  \right)
\end{equation}
\begin{equation}
\label{eq:10}
  Y^d(M_X) \sim
  \left(
    \begin{array}{ccc}
      \frac{8}{5}\delta\epsilon^5 &
      -\sqrt{2}\delta^2\epsilon^3 &
      \frac{4}{\sqrt{5}}\delta^2\epsilon \\
      \frac{2}{\sqrt{5}}\delta\epsilon^4 &
      \sqrt{\frac{2}{5}}\delta\epsilon^2 
      + \frac{16}{5\sqrt{5}}\delta^2\epsilon^2 &
      \sqrt{\frac{2}{5}}\delta^2 \\
      \frac{8}{5}\delta\epsilon^5 &
      \sqrt{2}\delta\epsilon^2 &
      r_b
    \end{array}
  \right)
\end{equation}
\begin{equation}
  \label{eq:11}
  Y^\nu(M_X) \sim
  \left(
    \begin{array}{ccc}
      \sqrt{2}\delta^3\epsilon^5 &
      2\delta\epsilon^3 &
      0 \\
      0 &
      \frac{6}{5\sqrt{5}}\delta^2\epsilon^2 &
      2\delta \\
      0 &
      \frac{6}{5}\delta^2\epsilon^2 &
      r_\nu
    \end{array}
  \right)
\end{equation}
\begin{equation}
  \label{eq:12}
  Y^e(M_X) \sim
  \left(
    \begin{array}{ccc}
      \frac{6}{5}\delta\epsilon^5 &
      0 &
      0 \\
      \frac{4}{\sqrt{5}}\delta\epsilon^4 &
      -3\sqrt{\frac{2}{5}}\delta\epsilon^2
      + \frac{12}{5\sqrt{5}}\delta^2\epsilon^2 &
      -3\sqrt{\frac{2}{5}}\delta^2 \\
      \frac{6}{5}\delta\epsilon^5 &
      \sqrt{2}\delta\epsilon^2 &
      1
    \end{array}
  \right)
\end{equation}

This model is consistent with all laboratory data; A global analysis 
\cite{Blazek:2003wz} assuming universal gaugino and sfermion masses but
non-universal Higgs masses, with $\tan\beta = 50$ is consistent with
sparticle and Higgs mass limits, fermion masses and mixing angles including
the LMA MSW, muon $g-2$, $b\rightarrow s\gamma$ and the LFV constraints.

Furthermore, in this model it is possible that the decay 
$B_s \rightarrow \mu^+\mu^-$ could occur at a rate \cite{Blazek:2003hv} 
close to the current limit of 
$BR(B_s \rightarrow \mu\mu) < 2.0 \times 10^{-6}$ which is much larger 
than the Standard Model prediction of 
$BR(B_s \rightarrow \mu\mu)\sim 1.5\times 10^{-10}$.

\begin{figure}[htp]
  \centering
  \epsfxsize=20pc
  \epsfbox{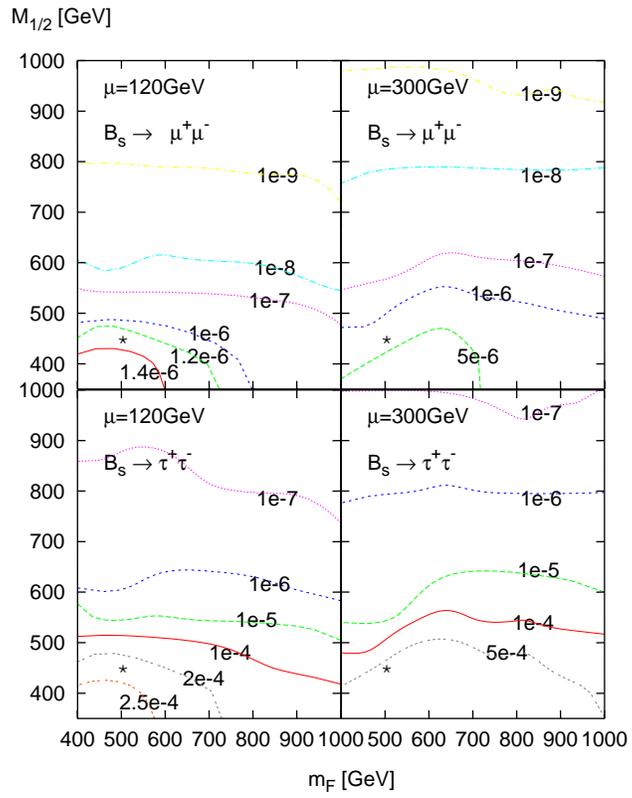}
  \caption{$B_s \rightarrow \mu^+\mu^-$ predictions}
    
  \label{fig:bs_graph}
\end{figure}

We now turn to understanding why the primordial off-diagonal terms are small.
This is entirely natural in the context of a type I string embedding. In fact,
a string construction which leads to a continuum of SUSY Pati-Salam models
is known \cite{Everett:2002pm} ( See also \cite{OtherTypeI} ). The idea here
is that everything is set up on two intersecting $D5$-branes. One of them,
the $5_1$-brane
has the gauge group $U(4)^{(1)}\otimes U(2)_L \otimes U(2)_R$, and the other,
the $5_2$-brane
has the gauge group $U(4)^{(2)}$. Higgs states are present which can break
$U(4)^{(1)}\otimes U(4)^{(2)} \rightarrow U(4)_{PS}$. Most of the $U(1)$s are
broken by the GS mechanism, but one remains: $U(1)_F$. 
 Furthermore, in this model $R_2 \ll R_1$, 
(``single brane limit''), and we have approximate gauge coupling unification.

The important point from a flavour point of view is that in this model the
Pati-Salam and Family breaking fields can have auxiliary fields which may
contribute to symmetry breaking; despite the fact that these are much smaller
than the dominant contributions coming from moduli, they can contribute
to the A-terms at the same order as the moduli, since \cite{Atermstuff}:
\begin{equation}
  \label{eq:14}
  A \supset F_\phi \partial_\phi \ln Y 
  = F_\phi \partial_\phi ln \phi^n = F_\phi \frac{n}{\Phi} \propto n m_{3/2}
\end{equation}

(The last step is since $F_\phi \propto m_{3/2} \phi$). Having done this, you
need a parameterisation for the auxiliary field VEVs. The usual one is that of
goldstino angles. In order to get the auxiliary field VEVs for the 
family group Higgs, it was necessary to pick a definite brane allocation. Picking
it to be an intersection state, the F-term VEVs and then the soft terms
can be written down \cite{King:2003kf}. Where $\sum_\alpha X_\alpha^2 = 1$:
\begin{eqnarray}
  \label{eq:15}
  m_F^2 = 
  \left(
    \begin{array}{ccc}
      a & 0 & 0 \\
      0 & a & 0 \\
      0 & 0 & b
    \end{array}
  \right)\;,\; a = 1 - \frac{3}{2}(X_S^2 + X_{T_3}^2)
  \;,\;
  b = 1 - 3 X_{T_3}^2 \\
  \label{eq:16}
  M_3 \approx M_2 \approx M_1 = \sqrt{3}m_{3/2}X_{T_1} \equiv M_{1/2} \\
  \label{eq:17}
  A = \sqrt{3}m_{3/2}
  \left(
    \begin{array}{ccc}
      d_1 + d_H + 5d_\Phi & d_1 + d_H + 3d_\Phi & d_2 + d_H + d_\Phi \\
      d_1 + d_H + 4d_\Phi & d_1 + d_H + 2d_\Phi & d_2 + d_H \\
      d_3 + d_H + 4d_\Phi & d_3 + d_H + 2d_\Phi & d_4 
    \end{array}
  \right)
\end{eqnarray}

Where, the $d$ parameters are:
\begin{eqnarray}
  \label{eq:18}
  d_1 = X_S - X_{T_1} - X_{T_2} &,&
  d_2 = \frac{1}{2}X_S - X_{T_1} - \frac{1}{2}X_{T_3} \\
  \label{eq:19}
  d_3 = \frac{1}{2}X_S - X_{T_1} - X_{T_2} + \frac{1}{2}X_{T_3}
  &,&
  d_4 = -X_{T_1}
\end{eqnarray}
\begin{equation}
  \label{eq:20}
  d_H = (S+S^*)^{\frac{1}{2}}X_H + (T_3+ T_3^*)^{1/2} X_{\overline{H}}
\end{equation}
\begin{equation}
  \label{eq:22}
  d_\Phi = (S+S^*)^\frac{1}{4}(T_3 + T_3^*)^{1/4} X_\Phi
\end{equation}

\begin{table}[htbp]
  \centering
  \begin{tabular}{|c|c|c|c|c|c|}
    \hline 
    Point & $X_S$ & $X_{T_i}$ & $X_\Phi$ & $X_H$ & $X_{\overline{H}}$ \\
    \hline
    A   & 0.50 & 0.50 & 0.00 & 0.00 & 0.00 \\
    B   & 0.54 & 0.49 & 0.00 & 0.00 & 0.00 \\
    C   & 0.27 & 0.27 & 0.84 & 0.00 & 0.00 \\
    D   & 0.27 & 0.27 & 0.00 & 0.60 & 0.60 \\
    \hline
  \end{tabular}
  \caption{Four benchmark points for the $U(1)_F$ model}
  \label{tab:benchmarks_u1}
\end{table}

\begin{figure}[htbp]
  \centering
  \input{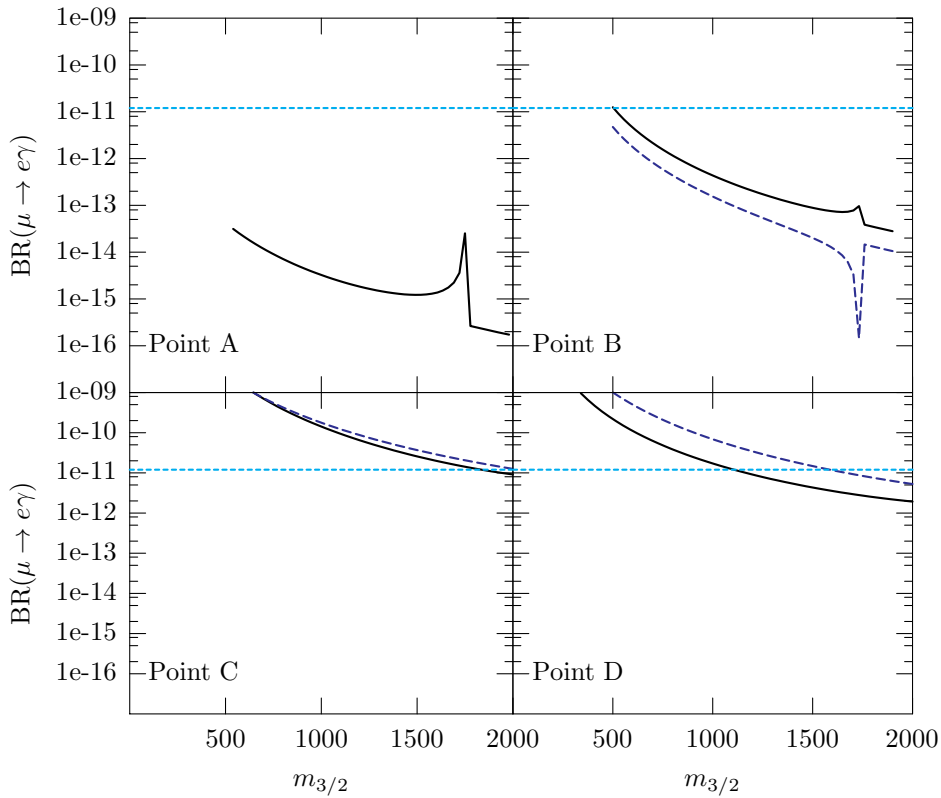}
  \caption{$\mu\rightarrow e\gamma$ in the $U(1)_F$ model. The dashed line
represents a model with no RH neutrino field, removing the see-saw contribution
to the process. The solid line represents `physics', with FV coming in general
from both primordial and see-saw sources. The horizontal line is the 2002
experimental limit. }
  \label{fig:meg_graph_u1}
\end{figure}

Having done this, we consider for benchmark points, A-D (given in
tab.~\ref{tab:benchmarks_u1}), which represent
the various independent sources of flavour violation. A represents no
primordial FV, B represents primordial FV from the moduli 
alone, C represents primordial FV from the Froggatt-Nielsen fields alone,
and D represents primordial FV from the Pati-Salam Higgs fields alone.
Fig.~\ref{fig:meg_graph_u1}
shows two lines for each benchmark point, one with the see-saw contribution
removed by removing the RH neutrino field, and one with a seesaw contribution.
We see that we are able to stay within the experimental limits 
\cite{Hagiwara:fs} at point C, and that the
see-saw source can {\em reduce} the amount of FV coming from the primordial
sector ( see fig.~\ref{fig:meg_graph_u1}, point D ).

We have assumed that the D-terms coming from $U(1)_F$ are small to keep D-term
contributions to LFV small; however with a similar setup, it is possible to suppress
LFV by having the D-terms very large \cite{Pokorski:talk}.

\section{$SU(3)$ family symmetry}
\label{sec:u3-family-symmetry}

We now look at a grand unified, non-abelian theory of flavour, 
$SO(10) \otimes SU(3)_F$. We also impose a 
$U(1) \otimes Z_2 \otimes \mathbf{R}$ global symmetry, which will
only allow certain Yukawa and $\mathcal{L}_{\mathrm{soft}}$ 
operators. Then we break $SO(10)$ first to the
Pati-Salam group and then to the MSSM group by two instances of Wilson line
breaking. 

Having done this, we break $SU(3)_F \rightarrow SU(2)_F$
with the field $\phi_3$, and then we break $SU(2)_F \rightarrow 0$ 
 with the field $\phi_{23}$:
\begin{equation}
  \label{eq:21}
  \left<\phi_3\right> = 
  \left(
    \begin{array}{c}
      0 \\
      0 \\
      1
    \end{array}
  \right) \otimes 
  \left(
    \begin{array}{cc}
      a_3^u & 0 \\
      0 & a_3^d 
    \end{array}
  \right)
  \;,\;
  \left<\phi_{23}
  \right>
  =
  \left(
    \begin{array}{c}
      0 \\
      1 \\
      1
    \end{array}
  \right) 
  b
\end{equation}

The charges under $U(1)_X$ restrict the form of the operators that form
the Yukawa elements and soft terms. Once you choose a set to match with the
Yukawa sector, then the operators which form the soft terms are completely
fixed. For example, one set of field assignments \cite{King:2003rf} allows
two (amongst many~) operators:
\begin{equation}
  \label{eq:23}
  \frac{1}{M^2} \psi\phi_3\psi\phi_3 h 
  \rightarrow
  \left(
    \begin{array}{ccc}
      0 & & \\
       & 0 & \\
       & & \overline{\epsilon}
    \end{array}
  \right)\;,\;
  \frac{\Sigma}{M^3}\psi\phi_{23}\psi\phi_{23}h
  \rightarrow
  \left(
    \begin{array}{ccc}
      0 & & \\
      & y\epsilon^2\overline{\epsilon} & y\epsilon^2\overline{\epsilon} \\
      & y\epsilon^2\overline{\epsilon} & y\epsilon^2\overline{\epsilon}
    \end{array}
  \right)
\end{equation}

There are operators which contribute to the Majorana matrix as well. In
the end, setting $\mathcal{O}(1)$ coefficients empirically, we get:
\begin{equation}
  \label{eq:24}
  Y^u \sim 
  \left(
    \begin{array}{ccc}
      0 & 1.2\epsilon^3 & 0.9\epsilon^3 \\
      -1.2\epsilon^3 & -\frac{2}{3}\epsilon^2 & -\frac{2}{3}\epsilon^2 \\
      -0.9\epsilon^3 &-\frac{2}{3}\epsilon^2 & 1
    \end{array}
  \right)\overline{\epsilon}\;,\;
  Y^d \sim
  \left(
    \begin{array}{ccc}
      0 & 1.6\overline{\epsilon}^3 & 0.7 \overline{\epsilon}^3 \\
      -1.6\overline{\epsilon^3} & \overline{\epsilon}^2 
      & \overline{\epsilon}^3 + \overline{\epsilon}^\frac{5}{2} \\
      -0.7\overline{\epsilon}^3 
      & \overline{\epsilon}^2 - \overline{\epsilon}^\frac{5}{2} & 1     
    \end{array}
  \right)\overline{\epsilon}
\end{equation}
\begin{equation}
  \label{eq:25}
  Y^\nu \sim 
  \left(
    \begin{array}{ccc}
      0 & 1.2\epsilon^2 & 0.9\epsilon^2 \\
      -1.2\epsilon^2 & -\alpha\epsilon^2 
      & -\alpha\epsilon^2 
      +\frac{\epsilon^3}{\sqrt{\overline{\epsilon}}} \\
      -0.9\epsilon^3 & -\alpha\epsilon^2 
      - \frac{\epsilon^3}{\sqrt{\overline{\epsilon}}} & 1
    \end{array}
  \right)\overline{\epsilon}
  \;,\;
  Y^e \sim
  \left(
    \begin{array}{ccc}
    0 & 1.6\overline{\epsilon}^3 & 0.7\overline{\epsilon}^3 \\
    -1.6\overline{\epsilon}^3 & 3\overline{\epsilon}^2 
    & 3\overline{\epsilon}^2 \\
    -0.7\overline{\epsilon}^3 & 3\overline{\epsilon}^2 & 1
\end{array}
  \right)\overline{\epsilon}
\end{equation}
\begin{equation}
  \label{eq:26}
  \frac{M_{RR}}{M_3} \sim
  \left(
    \begin{array}{ccc}
      \epsilon^6\overline{\epsilon}^3 & & \\
      & \epsilon^6\overline{\epsilon}^2 & \\
      & & 1
    \end{array}
  \right)
\end{equation}

The first RH neutrino dominates, and we predict 
$m_2/m_3 \sim \overline{\epsilon}$, $\tan\theta_{23} \sim 1.3$, 
$\tan\theta_{12} \sim 0.66$ and $\theta_{13} \sim \overline{\epsilon}$.

We also constrain the order of terms allowed in $\mathcal{L}_{\mathrm{soft}}$.
The scalar matrices come from $D$-terms, and the trilinears come from the
$\mathcal{F}$-terms \cite{Lsoftinsu3}:
\begin{equation}
  \label{eq:27}
  \mathcal{L}_{\mathrm{soft}} \supset
  \left\{
  \frac{S^\dag S \psi^\dag \psi }{M^2} 
  +
  \frac{S^\dag S \psi^\dag \phi^\dag_3 \psi \phi_£}{M^2} + \cdots 
  \right\}_D + \left\{ \frac{\cdots}{\cdots}\right\}_{\mathcal{F}}
\end{equation}
So, then the scalar matrices get terms like
\begin{equation}
  \label{eq:28}
  m^2_{\tilde\psi} \supset
  \left(
    \begin{array}{ccc}
      m^2_0 & & \\
      & m^2_0 & \\
      & & m^2_0 
    \end{array}
  \right)
  +
  \left(
    \begin{array}{ccc}
      0 & & \\
    & 0 & \\
    & &\overline{\epsilon}m^2_0
    \end{array}
  \right)
  + \cdots
\end{equation}

This leads to a characteristic pattern of SUSY masses, with FCNCs suppressed
by high powers of $\epsilon$ and $\overline{\epsilon}$.

\section{Conclusions}
\label{sec:conclusions}

A $U(1)$ family symmetry allows an understanding of fermion
masses and mixings, but does not address the SUSY flavour problem without
additional theoretical input. Such models are motivated by string
theories where $U(1)$s are abundant, and SUSY flavour changing may be 
controlled in a type I string embedding, even if the theory 
controls sparticle masses, dangerous new primordial flavour 
changing arise from Yukawa operators which lead to large off-diagonal 
soft trilinears.

A $SU(3)$ family symmetry allows (anti)symmetric Yukawa matrices, with SUSY
flavour changing controlled by the family symmetries. However, it is hard
to get non-abelian family symmetries from string theory.

\end{document}